\documentclass[9pt,letterpaper,twocolumn]{article}

\usepackage[latin1]{inputenc}
\usepackage{amsmath}
\usepackage{amsfonts}
\usepackage{amssymb}
\usepackage{braket}
\usepackage{enumerate}

\usepackage[left=1.00in, right=1.00in, top=1.00in, bottom=1.00in]{geometry}

\author{
Horace P. Yuen\\
Department of Electrical Engineering and Computer Science\\
Department of Physics and Astronomy\\
Northwestern University, Evanston Il. 60208\\
yuen@eecs.northwestern.edu }

\title{On the Foundations of Quantum Key Distribution --- \newline Reply to Renner and Beyond\thanks{}}

\begin{document}
\twocolumn[
\begin{@twocolumnfalse}
\maketitle
\begin{abstract}
\textit{In a recent note (arXiv:1209.2423) Renner claims that the
criticisms of Hirota and Yuen on the security foundation of
quantum key distribution arose from a logical mistake. In this
paper it is shown that Renner misrepresents the claims of Yuen and
also Hirota while adopting one main theorem of Yuen in lieu of his
own previous error. This leads to his incoherent position which
ignores quantitative security criterion levels that undermine the
current security claims, a main point of the Yuen and Hirota
criticisms. This security criterion issue has never been properly
addressed in the literature and is here fully discussed, as are
several common misconceptions on QKD security. Other foundational
issues are touched upon to bring out further the present
precarious  state of quantum key distribution security proofs.}
\newline
\newline
\end{abstract}
\end{@twocolumnfalse}
]

{
\renewcommand{\thefootnote}{\fnsymbol{footnote}}
\footnotetext[1]{This is a slightly revised version of
arXiv:1210.2804v1 and is identical to v2. It is going to appear in
Quantum ICT Research Institute Bulletin of Tamagawa University,
vol.3, no-1, p.1, 2013.} }
\section{INTRODUCTION}

In this paper we will respond to the recent Reply paper by Renner
[1] that the criticisms of Yuen [2-5]  and Hirota [6] on the
security of quantum key distribution (QKD) protocols are derived
from a logical error. While Hirota could speak for himself, some
related points in his paper would be included in our discussion.
Renner explicitly attributes an equivocal claim to us, and by an
incorrect argument in a footnote, claims to produce a
counter-example to our conclusion. In truth, the precise form of
our claim has been repeatedly given in [2-5]. Rather, Renner made
a fundamental error in [7-8] which has become the standard
interpretation of the trace distance criterion $d$ widely employed
in QKD. This incorrect interpretation leads to the current
prevalent QKD security claim that the generated key $K$ has a
probability $p \geq 1-d$ of being ideal [9-11]. In actuality, $K$
is not ideal with probability 1 for $d>0$ and may have a
probability $d$ of being found in total by an attacker Eve [2-5].
As brought out in detail in section III, the correct meaning of
$d$ gives a much weaker security guarantee than the wrong
interpretation in general. It is the consequence of this error in
concrete QKD protocols that Yuen and Hirota pointed out, which is
beyond rational dispute as will be shown in this paper.

Security is a quantitative issue. The exact level one has for a
given $l$-bit key $K$ is crucially important. In [1] $l$ is taken
to be $10^{6}$ and $d = 10^{-20}$. There are two sorts of
security, ``raw security'' [3] before $K$ is used and composition
security where Eve has additional information about it when $K$ is
used, for example from a known-plaintext attack. In raw security,
the ideal situation occurs when $K$ has the uniform distribution
$U$ to Eve. Since the earlier days of QKD [12], ``unconditional
security'' means the security result holds against all attacks
allowed by the laws of quantum physics, with quantitative
information theoretic security level that can be made arbitrarily
close to ideal through a \textit{security parameter}. If $d$ is
the maximum failure probability with ``failure'' meaning the key
is not ideal [7-11], security would be perfect with a large
probability $p \geq 1 - d$, but that is false. When $K$ has a
distribution $P$ to Eve, its quality is often measured by a
single-number security criterion, say the variational distance
$\delta(P, U)$ between $P$ and $U$. Since $\delta$ or $d$ is not a
bound on $1-p$, operational security meaning has to be given to
them through Eve's probabilities of success in estimating various
portions of $K$ and through Eve's average bit error rate (BER)
[2-5].

This paper would provide the details to elaborate on the
following:

\begin{enumerate}[\itshape(1)]
\item What Renner Claimed Before The Reply --- \newline The trace
distance $d$ is defined in (TD) of [1] with (TD) meaning $d \leq
\epsilon$, equation (1) of [1] says
\begin{equation}
\mathrm{(TD)} \rightarrow (\mathrm{UC} \hspace{2mm} secrecy)
\end{equation}
In [7-11] before this Reply paper [1], UC secrecy (of level at
least $\epsilon$, or ``$\epsilon$-secrecy'') means the generated
$K$ is ideal with probability $p \geq 1-\epsilon$. This is often
phrased in terms of the ``failure probability'' $1-p$ being less
than $\epsilon$. Thus, with $d$ interpreted as the maximum failure
probability [7-11], (1) is obtained to guarantee
$\epsilon$-secrecy when the level of $d$ is bounded by $\epsilon$.

\item What Yuen And Hirota Claimed --- \newline It was shown [2-5]
that Renner's interpretation of $d$ is incorrect and in fact $K$
is not uniform with probability 1 when $d > 0$, i.e., $p = 0$.
Furthermore, the levels of $d$ obtained in concrete protocols, in
theory [13] not to say in experiment [14], imply $K$ is very poor
compared to $U$ [2-6], for both raw and known-plaintext attack
security and for both Eve's sequence success probabilities and
BER.

\item What Renner Claimed In His Reply --- \newline The meaning of
(1) is now equivocal in [1]. In paragraph two, UC secrecy is still
claimed to be ``$\epsilon$-secret'' with a failure probability
$\leq \epsilon$, but the explanation of failure probability in
footnote [14] is given in terms of the correct sequence
probability meaning of $d$ first described in [5] but with no
reference. The BER meaning is not given. These two interpretations
of UC secrecy in [1] are contradictory, as indicated in point
\textit{(2)} above. By an arbitrary stipulation in footnote [15],
it is declared in [1] that $d = 10^{-20}$ for an $l=10^{6}$ bit
key is sufficiently secure. Together with distorting our correct
claim that the condition (HY) means the key is near-uniform to
that it is necessary for security, a ``logical error'' on Yuen and
Hirota is manufactured in [1] through a counter-example in
footnote [19]. This counter-example itself is infused with error
and confusion, including the \textit{same} conceptual confusion
that leads to the error described in \textit{(1)} above.

\item What Is Wrong With The Security Claim In [1] --- \newline In
addition to the above point \textit{(2)} there are fundamental
problems on the claims in [1] for $d=10^{-20}$ and $l=10^6$. Since
$K$ is then far from uniform, it cannot be used to subtract for
$leak_{EC}$, Eve's information gain from error correction, that is
employed in all recent security proofs. Also, why is such $d$
level ``sufficient'' for security? When $K$ is not near-uniform,
\textit{only} the users in a specific application can decide
whether a given $d$ level is sufficient. It \textit{cannot} be
prescribed in advance at $d=10^{-20}$. It is the responsibility of
the security analyst to spell out clearly the key rate and
security level tradeoff. Note that according to the most up to
date theoretical analysis of single-photon BB84 in [13],
$d=10^{-20}$ is nowhere to be found. Already in their presented
results the key rate is reduced to effectively zero at
$d=10^{-14}$, with a one-bit $K$ generated before message
authentication bits are accounted for.

\hspace{5mm} When the average guarantee in the security proofs is
converted to individual guarantee necessary for security claim on
an individual system, the level is reduced from $d$ to $d^{1/3}$
for Eve's sequence success probabilities [4]. Thus, $d=10^{-20}$
[1] reduces to $d^{1/3} > 10^{-7}$. For $d=10^{-14}$ [13],
$d^{1/3} > 10^{-5}$ and for  $d=10^{-6}$ [14],  $d^{1/3} =
10^{-2}$. These are poor to very poor security guarantees for any
application, and they remain so even under the wrong
interpretation. Such quantitative issues are among the main claims
of [2-6] not addressed in [1].

\item What Are The Other QKD Security Foundation Issues ---
\newline There are many other basic problems in the known QKD
security proofs that have been raised additionally in [15-19] but
not touched upon in [1] despite its title and references. There
are also several common but fundamental misconceptions in QKD
security that should be clarified. A most significant
misconception is that there is a security parameter in QKD
protocols that can bring security to an arbitrarily good level if
the key rate is below a certain threshold.
\end{enumerate}

In this paper, we will explain points \textit{(1)}-\textit{(5)} in
detail. In section II we will explain the above criterion issue to
settle the matter once for all. We will start by dispelling a
common misconception that QKD security is guaranteed by the laws
of quantum physics, either no-cloning or whatever Uncertainty
Relation. The \textit{necessary condition} for operational
quantitative security will be given. We will describe the severe
reduction of the guaranteed $d$ level to $d^{1/3}$, and the
importance of bringing Eve's BER on $K$ close to 1/2. While many
details on the points about $d$ itself can be found from [2-5], in
section III we will make just one basic point on the error of
interpreting $d$ as maximum failure probability, namely a new
fundamental argument on why the `proof' of such an interpretation
given in [7-8] involving a joint distribution is not only invalid
but is in fact irrelevant to the issue. All the security points
raised in [1] will be addressed. In section IV various security
proof issues concerning BB84 type protocols will be touched upon.
We will bring out the inevitable exchange of key rate and security
level in QKD systems, with the important consequence that there is
\textit{no} security parameter in QKD protocols that would render
it arbitrarily secure for a fixed key rate. We will point out the
incorrect step of subtracting $leak_{EC}$ to account for
information leak due to error correction. Some common
misconceptions about QKD security are summarized in section V.

The upshot is that the security foundation of QKD is indeed very
much shaken. General security cannot be established by experiments
and can only be proved theoretically. The present predicament is
that it is \textit{not} clear why and how a concrete QKD protocol
can be proved secure in principle.

\section{QKD SECURITY CRITERION AND NECESSARY SECURITY CONDITIONS}

In a QKD protocol of the BB84 type [20] two users A and B try to
establish a sequence of secret bits, the generated key $K$,
between themselves that no eavesdropper Eve can know even with any
active attack. The security is often claimed to be based on the
laws of quantum physics as if the latter have to be violated in
order for Eve to succeed. It is clear that quantum no-cloning is a
necessary but far from sufficient condition for security. In
particular, the possibility of approximate cloning shows the issue
is a more complicated quantitative one. The more prevalent
intuitive security idea is quantum disturbance-information
trade-off, that the users could tell the presence of Eve by
monitoring the system disturbance level if she gains an amount of
``information'' on $K$ exceeding a given design level. Indeed,
intrusion level estimation is a key part of all the typical QKD
approaches. Henceforth the term QKD is used with the understanding
that intrusion level estimation is involved.

To get sizable disturbance relative to the signal that can be
readily estimated in QKD, the signal level needs to be low, say a
single photon in BB84. Thus, the disturbance induced by Eve is
easily masked by other unavoidable disturbance in a concrete
realistic system even when such imperfection is small for other
purposes. Furthermore, in an active attack Eve could in principle
transform the quantum signals in many different ways and the users
have to estimate her information gain under a given level of
tolerable disturbance. It is now clear that security is a
\textit{quantitative} and \textit{complicated} matter, and that
there is no simple intuitive reason why any net key bits can be
generated in QKD with whatever security, especially when the bits
used for message authentication necessary for defending against
man-in-the-middle attack are counted.

What security criterion should one use to measure the quantitative
security level and why? In the literature this issue has
\textit{never} been correctly addressed. The mutual (accessible)
information was used from the beginning but was found to contain a
major loophole [21,22] and is by now largely abandoned. The trace
distance criterion $d$ [23,7-8] is at present nearly universally
employed in QKD security analysis which is cited in [1] as the
criterion that leads to ``UC secrecy''.

What is the level of $d$ needed for UC secrecy? While one can
distinguish perfect secrecy from UC secrecy, adequate UC security
cannot be established by mere terminology or definition. It
appears that the QKD security criterion is often thought to be a
matter of choice by the designer, a wrong conception as we show
presently. In [5] the following criteria are given in terms of
Eve's optimal probabilities $p_1$ of successfully estimating
various subsets of $K$ from her attack. For raw security [3] where
Eve only has information from the key generation process, the
conditions are, with $K^*$ being any subset of $K$ and for any
value $k^*$ of $K^*$,
\begin{equation}
p_{1}(k^{*}) \leq 2^{-|K^{*}|}+\epsilon'
\end{equation}
for some chosen level $\epsilon'$ [5]. Under known-plaintext
attack where Eve knows a subset segment $K_1 = k_1$ of $K$ and
estimates a subset $K_{2}^{*}$ in the rest of $K$, the condition
is, for some level of $\epsilon''$,
\begin{equation}
p_{1}(k_{2}^{*}|K_{1}=k_{1}) \leq 2^{-|K_{2}^{*}|}+\epsilon''
\end{equation}
These probabilities have direct operational meaning in contrast to
theoretical entities such as $d$ or mutual information. The users
have to decide what the $\epsilon'$ and $\epsilon''$ are for the
cryptosystem to be sufficiently secure operationally in a
particular application. In particular, if these levels cannot be
guaranteed it means Eve may be able to guess the key portion $K^*$
or $K_{2}^{*}$ with a probability exceeding the prescribed level
chosen by the users, thus the cryptosystem is \textit{not} proven
secure to its operational specification! Hence (2)-(3) are
\textit{necessary conditions} for security. They are not
sufficient for one-time pad use of $K$, as discussed later.

Among different composition security situations, known-plaintext
attacks have to be included in QKD security proofs. As discussed
in [3], the raw security of conventional symmetric-key ciphers is
far better than that of concrete QKD systems.

As explained in [2], Eve derives from her probe measurement a
whole distribution $P$ on all the $2^l$ possible $K$ values. A
single-number criterion merely expresses a constraint on $P$, but
$P$ itself should be compared to $U$ for operational security
guarantees. In particular, one has the form given in the left
sides of (2)-(3) above for Eve's sequence success probabilities.
In the ideal case, $\epsilon' = \epsilon'' = 0$ in (2)-(3). The
levels $\epsilon'$ and $\epsilon''$ can be stipulated by the
system designer for different security needs. Under a $d \leq
\epsilon$ guarantee, (2)-(3) hold only when averaged over all
relevant key values [5] with $\epsilon' = \epsilon'' = \epsilon$.

From Markov inequality [24] such average guarantee can be
converted into the individual guarantees (2)-(3) for
\textit{proper comparison with $U$} [25]. Operationally, average
guarantee is \textit{not} sufficient also because ``failure
probability'' of some sort is required in the quality control of
individual items in any production system. Thus, we have (2)-(3)
with
\begin{equation}
\epsilon' = \epsilon'' = d^{1/3}
\end{equation}
due to averaging of $d$ with respect to the possible $K$ values
and the privacy amplification codes given in security proofs
[4,5].

Our averaged conditions [5] are obtained for the classical
variational distance [24] which is bounded by $d$ upon measurement
from Eve. They do not seem to have appeared before [26] in either
the classical or quantum literature other than deterministic bit
leak in raw security brought up in [3]. Probabilistic bit leaks of
any level are covered in (2)-(3), and such leaks must also be
guaranteed by quantitative bounds. Note that equality can be
achieved for these bounds, i.e., there are Eve's distributions on
$K$ compatible with the $d \leq \epsilon$ guarantee which satisfy
(2)-(3) with equality [2-5]. This shows they can be used with
equality to measure the quantitative security guarantee on $K$.

What would be a sufficient condition for security? If $\epsilon$'
and $\epsilon$'' are not small in the right scale with respect to
$l$, (2)-(3) may \textit{not} be sufficient depending on the
application. Recall that the comparison reference of the
distribution $P$ of $K$ is $U$. When $K$ is used in one-time pad
form, in addition to (2)-(3) Eve's average BER $p_b$ in her
estimate of the $K$ bits has to be close to 1/2 for security.
(Note that $p_b$ accounts for the correlation between the bits in
$K$ from its definition [4].) This is well known in data
communications and is easily seen, that an incorrect sequence
estimate on $K$ may nevertheless produce a preponderance of
correctlyy estimated key bits similar to what one may get from a
biased a priori distribution of $K$ that is different from $U$. It
turns out that [4] only
\begin{equation}
\frac{1}{2} - p_b \leq d^\frac{1}{4} /2 \sqrt{\log_2 e}
\end{equation}
can be guaranteed for the whole $K$ in raw security, there is no
subset guarantee for either raw or known-plaintext attack
security. However, if $d \sim 2^{-l}$ for $l \gg 1$ so that $K$ is
near-uniform, it appears $K$ should be quantitatively secure for
all conceivable applications as stated in [15]. Note that no
composition security argument from the mere form of $d$ [23] can
guarantee $p_b$ under known-plaintext attacks [4], while the wrong
interpretation can [11], because $K$ is $U$ with a high
probability $p \geq 1-d$.

\section{THE INCORRECT INTERPRETATION OF $d$ AND CLASSICAL CRYPTOGRAPHY}

The prevalent interpretation is that $d$ gives the probability
that $K$ is different from $U$ with Eve's probe disconnected from
$K$ and thus giving composition security also [7-11]. This
interpretation has repeatedly been pointed out to be incorrect in
[2-5] to no avail, until the appearance of [1], which no longer
cites such an interpretation but instead the correct one! The
origin of the error comes from the interpretation of the
variational distance $\delta(P,Q)$,
\begin{equation}
\delta(P,Q)=\frac{1}{2}\sum_{i}|P_i-Q_i|
\end{equation} between two classical probability distributions $P$ and $Q$ which is given to Proposition 2.1.1 in [7], that ``the two settings described by $P$ and $P'$, respectively, cannot differ with probability more than $\epsilon$.'' In our present notation or that of [8], $P' = Q$, and $d$ is interpreted equivalently from Lemma 1 of [8] as the ``probability that two random experiments described by $P$ and $Q$, respectively, are different''. We would not repeat the reasons and simple counter-examples [2-5] on why this interpretation is wrong. It does not follow from the mathematical statement of his Proposition 2.1.1, or the equivalent Lemma 1 of [8], through a joint distribution which gives $P$ and $Q$ as marginals, but rather from conceptual and verbal confusions. Instead, we point out here that any such joint distribution is irrelevant to the meaning of $\delta(P,Q)$. This is simply because the marginal distribution is just $P$ regardless of what the underlying space of $P$ is joined to. $P$ does not suddenly become $Q$ with a probablity $\delta(P,Q)$ in the presence of the given joint distribution. The wrong interpretation arose from basic conceptual confusions about the relation of probability concepts to the real world. It is amazing that it has perpetuated as far and as long as it has.

The variational distance is a well studied concept and nowhere
else could one find such a strong interpretation as given in
[7-8]. In particular, $d$ is not so interpreted in [23]. Indeed,
it is shown in [3] and easily seen from (6) that when $d > 0$, the
distribution of $K$ is \textit{not} $U$ with probability 1 (no
probability issue here really) instead of $d$. Subtle and
equivocal words in [1] may suggest that the wrong and correct
interpretations of $d$ (equivalently $\delta$) are similar.
Although the two interpretations quantitatively contradict each
other, one may perhaps think they are numerically close. In
particular, since ``failure'' includes the event where the whole
$K$ is compromised, it is important to understand the difference
between the two interpretations precisely, as follows.

Prior to ref [5], which correctly proves known-plaintext attack
security under $d \leq \epsilon$ for the first time, in the
literature there are two incorrect/incomplete proofs of universal
composition security. One of them [11] is invalid since it
utilizes the wrong interpretation of $d$. With (3) from [5],
known-plaintext attack security is established for Eve's sequence
success probabilities but there is \textit{no} similar guarantee
for Eve's BER. In contrast, under the wrong interpretation Eve's
BER $p_b = \frac{1}{2}$ with a probability $\geq 1-d$ for every
$k$, on which counter-examples are easily constructed. In general,
each different composition situation has to be treated under the
correct meaning of $d$ for quantitative guarantee, which cannot be
given by just $d$ or $\delta$ since they are not operational
criteria. This fact alone shows the composition security claim on
$d$ in [23] in incomplete or invalid, since mathematical
representation of operation security is lacking.

A further difference is that if $K$ is not at least near uniform,
one cannot use it to subtract for the bits $leak_{EC}$, given by
(8) in section IV, while such bits need to be used in the middle
of a valid security proof. Another difference is that Markov
inequality needs to be applied only once under the wrong
interpretation since there is no $K$-average needed, which results
in $d^{1/2}$ instead of $d^{1/3}$ in (4).

Even \textit{assuming} the wrong interpretation is true, the
relatively large value of $d$ that can be obtained is quite
worrysome. For $d=10^{-20}$, the operational guarantee (2)-(3) for
a $10^6$ bit key is not better than that of a 66 bit key! An
arbitrary reason of system imperfection level given in footnote
[15] of [1] is used to justify such numerical values. But why is
$d = 10^{-20}$ sufficient for UC secrecy? In fact, the raw
security operational guarantee (1)-(2) for $d = 10^{-20}$ is much
worse than that obtained in conventional symmetric key ciphers
[3].

Furthermore, there is no hint that such a $d$ level of $10^{-20}$
can be obtained in a concrete protocol.  If one takes into account
Markov inequality for individual guarantee as discussed in section
II, only an effective $d^{1/3} > 10^{-7}$ is obtained for $d =
10^{-20}$ after the $K$ value average and privacy amplification
code average are accounted for [4]. The effective $d^{1/3}$ value
of $> 10^{-7}$ for $d=10^{-20}$ is already very large for $l =
10^3$, not to say $l=10^6$. The \textit{only} concrete
experimental protocol with quantified security level is given in
[14,27] with effectively $d = 10^{-6}$. Then $d^{1/3}=10^{-2}$
from [14] may entail a very drastic breach of security. Note that
the $d=10^{-20}$ level cannot even be achieved for a positive key
rate in a ``tight finite-key'' analysis of single-photon BB84
[13], for which the best $d=10^{-14}$ is obtained for $l=1$! It
should be emphasized that these effective $d^{1/3}$ values give
\textit{poor} security guarantee even according to the wrong
interpretation. The corresponding BER guarantee of (5) is
similarly poor.

In this connection, it is important to note that the size of $d$
should be measured with respect to $2^{-l}$ according to the
correct interpretation (2)-(3), not with respect to 1 according to
the incorrect interpretation. This has been a major source of
confusion, that since the system is evidently secure or ideal when
a criterion takes the value zero hence it should be secure for a
small value of the criterion. Yes, this is correct if
``smallness'' is measured in the correct scale, but 1 is not
always the scale, an elementary point that is often forgotten when
relative dimensional measure is ignored.

Similarly, the criterion $d$ as ``distinguishability advantage''
is used to justify $d$ as a security criterion in [23], which is
also the justification for using variational distance in some
classical cryptography work brought up in the last paragraph of
[1]. While the distinguishability advantage was only established
for binary decisions, it is now established [5] for N-ary
decisions for N between 2 and $2^l$. However, the relevant point
in this connection is that the required level $\epsilon$ in $d
\leq \epsilon$ depends on what N in the N-ary decision is. A value
good compared to $\frac{1}{2}$ for N=2 may be very inadequate
relative to 1/N for N = $2^l$, as we just discussed. This N-ary
issue is another reason why composition security proof has to be
spelled out precisely and quantitatively. Security is a
quantitative issue through and through. Further discussion of such
$d$ meaning is given in [4].

The quantitative counter-example in [1] is irrelevant to begin
with since we never deny (1) in its correct sense and we only
insist (HY) is necessary for a near-uniform $K$ when $l$ is large.
It may be mentioned that the counter-example uses a very strict
meaning for his vague condition (HY) that neither Yuen nor Hirota
ever indicated. The construction in the counter-example betrays
the \textit{same confusion} which underlies the erroneous
interpretation of $d$ [7-8]. In the counter-example, $\delta$ or
$d$ or $\epsilon$ is fixed at $2^{-l}$ and there is \textit{no
room} for another $\epsilon = 10^{-20}$ ``by construction''! This
is one conspicuous example of the several incoherences in [1].

In classical cryptography practice, encryption security is based
on complexity, search for known-plaintext attacks on symmetric key
ciphers and other computational ones in asymmetric key ciphers.
The information theoretic security we talk about here for QKD
plays no role except for one-time pad. Thus, the claim of [1] that
classical cryptography is compromised without a small enough $d$
is false, for this and the following reasons.

The bound storage model [28] with controllable information
theoretic security is not used in practice while it has a
criterion related to $d$, but there is a security parameter in
[28] that could make it arbitrarily small which is not available
in QKD. In particular, the key length $l$ itself is not such a
parameter once the proper criterion is employed in QKD [2], a
point that will be elaborated in the next section IV. On the other
hand, security is not fully established in [28] unless the
criterion value goes to zero, precisely because N-ary decisions as
well as Eve's bit error rate are not treated. In fact, security
under known-plaintext attacks, which is the real issue for
symmetric key ciphers [3], is also not treated in [28].

In public key cryptography the variational distance criterion from
complexity consideration plays no role in practice. In fact the
probabilistic encryption schemes that utilize such theory is not
used due to its slow speed. Similar to [28], security for public
key is not established in principle for N-ary decisions, Eve's bit
error rate, and for known-plaintext attacks.

The actual situation is that other than one-time pad, no protocol
in classical cryptography has been proven secure, information
theoretically or computationally. Cryptography is still very much
an art. Quantum cryptography aspires to provable security, a lofty
goal that has been repeatedly claimed to be achieved from numerous
errors of reasoning. Since security is a serious matter and cannot
be established experimentally, we should examine all the security
proof steps more carefully. A concise discussion of such steps and
the state of QKD security proofs is given next.

\section{QKD   SECURITY   PROOF   STATUS}

There are five main steps involved in the general security proof
of a BB84-type QKD protocol, assuming the physical modelling is
complete and correct:
\begin{enumerate}[(i)]
\item Pick a security criterion and establish its operational
guarantee is adequate; \item Measure the quantum bit error rate
(QBER) on the checked qubits and transfer it with proper
statistical margin to the sifted key $K''$; \item Bound Eve's
relevant information on $K''$ under an arbitrary joint attack;
\item Apply an open error correcting code (EEC) and bound Eve's
information on the corrected key $K'$; \item Apply an open privacy
amplification code (PAC) to generate the final key $K$ and bound
Eve's information on $K$ according to the chosen criterion to
obtain its quantitative level of security.
\end{enumerate}

Each of these five steps has been treated incorrectly since the
early days of QKD security proofs. At present, step (i) is almost
resolved (apart from Eve's general bit error rate) in one way
through the criterion $d$ via (2)-(5) above. Step (v) can be
resolved by the classical Leftover Hash Lemma [29].  We will
discuss the other three steps in turn, the main impediment to
progress in security proof is from steps (iii) and (iv).

Historically the Shor-Preskill proof [30] is most influential and
widely quoted, but it is incomplete/incorrect for all five steps.
Here it will be used as a representative and the other security
approaches and proofs other than [13] will not be discussed. The
Shor-Preskill proof employs the mutual accessible information
criterion $I_a$ without insisting it be small enough. (In contrast
to the impression from [21,22], the $I_a$ criterion is actually
fine if its level is at or below $2^{-l}$ for an $l$-bit key $K$
[31].)  The transfer of QBER is later amended in [32] for general
joint attacks, which is still incorrect because it involves
classical counting instead of qubit counting. It appears that
correct quantum counting can be developed [33], which gives wider
fluctuation or lower security level with a factor of two reduction
in the exponent.

The major difficulty in QKD security proof arises from the
correlation between key bits that are introduced by Eve's active
joint attack and the user's ECC and PAC. To account for such
correlation from a joint attack, step (iii) has mostly been
achieved by some sort of symmetrization which does not appear to
be valid. How does one get symmetry from an asymmetric situation?
The usual argument (see, for example, the reduction of a general
attack to collective attack in [7]) involving an openly known
permutation cannot do any work since Eve knows it and could just
rearrange back. A new argument is used in [13] which involves
incorrect classical counting on qubits similar to [32] and
moreover, does not work for sufficiently small $d$ [15].

The information Eve has on the chosen ECC and PAC are not
accounted for in the Shor-Preskill proof. In a direct development
of the Shor-Preskill approach, Hayashi has recently incorporated
such information for ECC [34] and PAC [35], which are yet to be
evaluated for concrete protocols under general attack. In the
meantime, the ECC information leak expression
\begin{equation}
leak_{EC}=h(\mathrm{QBER})
\end{equation}
where $h(\cdot)$ is the binary entropy function, is employed by
him [36] and in fact universally [9,13,37] to account for such
leak. It is pointed out [15] that there is the possibility of
information leak from ECC similar to quantum information locking
leak [15] that undermines inadequate values of accessible
information as a security criterion, and which is neglected in the
expression (7). Furthermore, (7) can be justified only for
collective attacks asymptotically. Collective attack is extremely
restrictive, Eve can launch what is called a joint attack without
any entanglement by just attacking a portion of the key bits
(which seems to suggest already that collective attacks cannot be
optimal for any of Eve's aim, not to mention for this $leak_{EC}$
issue). Indeed, \textit{no} justification for such a crucial
treatment of step (iv) by (7) has ever been spelled out because
there is \textit{none}. It cannot be true for all attacks if one
examines its meaning [15]. This ECC information leakage problem
(iv) and also the joint attack problem (iii) appear to be very
difficult to resolve in QKD security proofs.

The condition (7) by itself shows that the near-universal step of
subtracting it from the generated key bits to get the final $K$ is
\textit{invalid}, unless perhaps when the $d$ level of $K$ is so
small that (2)-(3) imply the bits are nearly uniform and $K$
functions effectively as $U$. This is a problem \textit{even} if
the users decide that a given large $d$ level is sufficient for
security. The security proof itself is supposedly carried out with
uniform bits in the amount (7). Note that Even could launch a
joint attack just to invalidate (7) regardless of whether
collective attack is optimum from the viewpoint of her information
gain on $K$. She may want to minimize the users' key rate which
may not turn out positive.

Apart from all these theory problems, the security proof claims
are often used by experimentalists to claim security for their
systems in an invalid way. For example, the Shor-Preskill
asymptotic key rate is often quoted as the system capability, with
\textit{no} mention of the criterion and its quantitative level.
Equally significantly, Shor-Preskill only claimed to have
established such rate for a joint CSS code as ECC and PAC. In
[38], for example, the cascade reconciliation protocol is used for
error correction which has numerous problems [39] and universal
hashing is used for PAC. However, it has \textit{never} been shown
that the Shor-Preskill key rate applies to such error correction
and privacy amplification procedures.

The asymptotic convergence rate for various criteria yields the
actual (asymptotic) key rate for fixed levels of $d$ or $p_1$
[2,25], and is not given in [24] for its mutual information
criterion. In this connection, we would like to bring out a
\textit{common misconception} concerning QKD security. Since [30]
it is often thought that as long as the key rate is below a
certain threshold, security level can be made arbitrarily close to
the ideal when the key length $l$ is indefinitely increased. That
is, $l$ is taken to be a security parameter, and that is likely
why only the secure key rate is quoted in many papers including
[38]. Perhaps this is thought to be in analogy with Shannon's
Channel Coding Theorem [24], which says that for data rate below
capacity, the error rate can be made arbitrarily small for long
enough block length. Sometimes it is thought that finite privacy
amplification is what renders this untrue. We would like to point
out here that the problem is present even asymptotically for any
$l \rightarrow \infty$, as follows [2].

For key rate below a threshold, let us assume it is indeed proved
that Eve's accessible information $I_a$ (or $d$) goes to 0 as $l
\rightarrow \infty$, exponentially as $\sim 2^{-\lambda l}$ for
some $0<\lambda<1$,
\begin{equation}
d \sim 2^{-\lambda l} \mathrm{ \hspace{2mm} or \hspace{2mm} }
I_a/l \sim 2^{-\lambda l}
\end{equation}
The situation for finite $l$ is the same. The security level for
those $l$ bits is very different depending on what exactly
$\lambda$ is. It is near ideal for $\lambda = 1$ but very far from
ideal for $\lambda << 1$. Indeed, Eve's maximum probability $p_1$
of estimating the whole $K$ sets the limit on the number of
uniform bits that can be generated since $p_1=2^{-n}$ for $n$
uniformly distributed bits. Thus, it is the \textit{rate} of $p_1$
or equivalently $I_a/l$ going to zero that determines the rate of
uniform key generation, \textit{not} the original key rate
threshold [2,15]. It turns out the convergence rate $\lambda$ in
(8) is very small for $d$ in [13], and not evaluated for $I_a/l$
in other proofs except [27] which leads to an even smaller
$\lambda$ [14]. With $d=10^{-20}$ and $l=10^6$, $\lambda \sim
\frac{2}{3}\times 10^{-4}$ resulting in 66 bits guarantee of
(2)-(3) for $10^6$ bits, or just 22 bits from (4) after Markov
inequality is applied. In [13] the best $d=10^{-14}$ or $d^{1/3} >
10^{-5}$, and in [14] $I_a/l \sim 10^{-6}$ equivalent to $d^{1/3}
\sim 10^{-2}$.

One can relax uniform $K$ to $\epsilon$-secrecy via
$\epsilon$-smooth entropy [40]. Intuitively, one cannot expect
much would be accomplished when $\epsilon$ is only moderately
larger than $2^{-l}$. In fact, even for very large $d$ for a given
$l$, the results of [13] shows the key rate is still very low.

Thus, the exchange of key rate and security level is a
\textit{fundamental} fact in all QKD protocols, asymptotic or
finite, and $l$ is not a security parameter. In fact, one needs to
prove that a \textit{positive} exponent $\lambda >0$ would result
in (8) which is far from guaranteed. This is especially the case
when all system imperfections and message authentication bits are
taken into account. Together with the numerical values obtained in
[13], this fundamental tradeoff between key rate and security
level gives a grim picture of the usefulness of BB84 type
protocols.

In QKD security proofs there are numerous problems associated with
physical modelling that have been ignored or neglected. We may
point out the case of general lossy channel security [16], photon
number splitting attacks on multi-photon sources and decoy states
[18], and heterodyne-resend attack in CV-QKD [19]. Security is
seriously undermined in the last two situations against the
prevalent security claims on them. In particular, a grave issue
that has been generally overlooked is to what extent the users
could accurately determine the various system parameters such as
loss, a serious \textit{robustness} issue for security. The well
known detector blinding attacks [41] shows detailed detector
behavior has to be explicitly represented in a real security proof
[17], but so far it has not been done.

\section {COMMON MISCONCEPTIONS ON QKD SECURITY}

The list in the following corrects some major misconceptions on
QKD security, most of which have been discussed in this paper as
part of our response to [1].
\begin{enumerate}[(a)]
\item Any single-number security criterion, other than the wrong
interpretation of $d$ in [7,8], is not sufficient for security by
itself. For operationally meaningful security guarantee, it has to
be quantitatively reduced to bounds on Eve's various success
probabilities in estimating segments of the key and also her
average bit error rate.

\item One cannot prescribe, as done in [1], that some chosen
numerical level of a criterion is always sufficient for security
when the level is far from ideal. It is the application user of
the cryptosystem who decides what level is adequate for a specific
application.

\item There is a fundamental exchange between key rate and
security level. It is not the case that security can be made
arbitrarily close to ideal for key rate below a certain threshold.
It is the cryptosystem designer's responsibility to evaluate such
quantitative tradeoff. The results of [13] give poor security
level even at very low key rate.

\item Contrary to widespread impression, there is no valid QKD
general security proof in the literature. For example, the error
correction step has never been treated correctly. The burden of
proof is on those who claim security, not on other to produce a
specific counter-example on the security claim.

\item As a consequence of (d) and in view of the fundamental
difficulties discussed in this paper, QKD is at present no
different in security status from other cryptosystems under study
or in use. It does not have the advantage of having been proved
unconditionally secure in principle.

\item The problem of complete system representation for security
claim is not a ``practical security'' issue for the application
user, but rather a basic one. The incomplete modelling of system
component behavior, such as photodetector temporal response to
different input signal levels, is not a mere ``side channel''
issue but a main issue of model completeness, without which there
can be no proof of security.
\end{enumerate}

\section{CONCLUSION}

It is hard to avoid the impression that Eve's standpoint has
rarely been taken seriously in the literature and the main concern
has been to claim security. A common mistake in general security
proofs is to analyze only one type of attacks but claiming
unconditional security against all possible attacks. Security is a
serious matter. There are an unlimited number of attack scenarios,
thus security can only be established theoretically if at all and
the burden of proof is on those who claim security. Attacks from
the Norway group [41] shows how dangerous a faulty claim may be,
with security totally compromised in an unexpected way, a
situation actually familiar in conventional cryptography. When
addressing security issues it would be good to keep the following
question in mind:

How did we come to the present QKD security predicament with
endless invalid security proofs?

\section*{ACKNOWLEDGEMENT}

I would like to thank Greg Kanter for useful discussions. My work
referred to in this paper was supported by the Air Force Office of
Scientific Research and the Defense Advanced Research Project
Agency.


\begin{thebibliography}{9}
\bibitem{ref25} These individual guarantee conditions are cited in footnote [14] of [1] as the significance of $d$ without the averaging, though averaging is complained about in footnote [18] of [1].
\bibitem{ref26} Thus, our condition (2)-(3) is just appropriated in footnote [14] of [1], apparently because it gives the correct operational significance of $d$ while [7-11] give an incorrect one.
\end{thebibliography}
\end{document}